\documentclass[a4paper,twoside]{article}
\newcommand{\affil}[1]{$^{\rm #1}$}

\usepackage[authoryear]{natbib}
\bibpunct{(}{)}{;}{a}{}{,}
\usepackage{graphicx}
\date{} %Please leave the date blank
\newcommand{\co}{$^{12}$CO(1-0)}

\newcommand{\mh}{H$_{2}$}

\newcommand{\hi}{H\,{\sc i}}
\newcommand{\ha}{H{$\alpha$}}

\newcommand{\cden}{cm$^{-2}$}
\newcommand{\vden}{cm$^{-3}$}

\newcommand{\msun}{M$_{\odot}$}
\newcommand{\kms}{km~s$^{-1}$}

\newcommand{\hcop}{HCO$^{+}$}

\newcommand{\xco}{$X_{\rm CO}$}

\title{\large\bf\flushleft The Molecular Ridge Close to 30\,Doradus in the Large Magellanic Cloud}
\author{\parbox{\textwidth}{\flushleft
\vspace{-0.5cm}
{\it J\"urgen Ott\affil{A,B,L}, Tony Wong\affil{C}, Jorge L. Pineda\affil{D}, Annie Hughes\affil{E,F}, Erik Muller\affil{F,M}, Zhi-Yun Li\affil{G}, Min Wang\affil{H}, Lister Staveley-Smith\affil{I}, Yasuo Fukui\affil{J}, Axel Wei{\ss}\affil{K}, Christian Henkel\affil{K}, \& Ulrich Klein\affil{D}}\\
\vspace{0.4cm}
{\small \affil{A}\,National Radio Astronomy Observatory, 520 Edgemont Road, Charlottesville, VA, 22903, USA}\\
{\small \affil{B}\,California Institute of Technology, 1200 E. California Blvd., Caltech Astronomy 104--25, Pasadena, CA, 91125--2400, USA}\\
{\small \affil{C}\,Department of Astronomy, University of Illinois, 1002 W. Green St., Urbana, IL 61801, USA}\\
{\small \affil{D}\,Argelander Institut f\"ur Astronomie, Universit\"at Bonn, Auf dem H\"ugel 71, 53121 Bonn, Germany}\\
{\small \affil{E}\,Centre for Supercomputing and Astrophysics, Swinburne University of Technology, Hawthorn VIC 3122, Australia}\\
{\small \affil{F}\,CSIRO Australia Telescope National Facility, Cnr Vimiera \& Pembroke Roads, Marsfield, NSW 2122, Australia}\\
{\small \affil{G}\,Department of Astronomy, University of Virginia, PO Box 400325, Charlottesville, VA 22903-4325, USA}\\
{\small \affil{H}\,Purple Mountain Observatories, CAS, 2 West Beijing Road, Nanjing 210008, China}\\
{\small \affil{I}\,School of Physics M013, University of Western Australia, Crawley WA 6009, Australia}\\
{\small \affil{J}\,Department of Astrophysics, Nagoya University, Furocho, Chikusaku, Nagoya 464-8602, Japan}\\
{\small \affil{K}\,Max-Planck-Institut f\"ur Radioastronomie, Auf dem H\"ugel 69, 53121 Bonn, Germany}\\
{\small \affil{L}\,J\"urgen Ott is a Jansky Fellow of the National Radio Astronomy Observatory. E-mail: jott@nrao.edu}\\
{\small \affil{M}\,Bolton Fellow}}}
\begin{document}
\twocolumn[
\begin{changemargin}{.8cm}{.5cm}
\begin{minipage}{.9\textwidth}
\vspace{-1cm}
\maketitle
%
%
%%%%%%%%%%%%%     ABSTRACT    %%%%%%%%%%%%%
%Abstract of no more than 200 words here.
\small{\bf Abstract:} 

With the ATNF Mopra telescope we are performing a survey in the \co\
line to map the molecular gas in the Large Magellanic Cloud (LMC). For
some regions we also obtained interferometric maps of the high density
gas tracers \hcop\ and HCN with the Australia Telescope Compact Array
(ATCA). Here we discuss the properties of the elongated molecular
complex that stretches about 2\,kpc southward from 30\,Doradus. Our
data suggests that the complex, which we refer to as the ``molecular
ridge,'' is not a coherent feature but consists of many smaller clumps
that share the same formation history. Likely molecular cloud
formation triggers are shocks and shearing forces that are present in
the surrounding south-eastern \hi\ overdensity region, a region
influenced by strong ram pressure and tidal forces. The molecular
ridge is at the western edge of the the overdensity region where a
bifurcated velocity structure transitions into a single disk velocity
component. We find that the \co\ and \hi\ emission peaks in the
molecular ridge are typically near each other but never coincide. A
likely explanation is the conversion of warmer, low-opacity \hi\ to
colder, high-opacity \hi\ from which \mh\ subsequently forms. On
smaller scales, we find that very dense molecular gas, as traced by
interferometric HCO$^{+}$ and HCN maps, is associated with star
formation along shocked filaments and with rims of expanding
shell-like structures, both created by feedback from massive stars.

%%%%%%%%%%%%%     KEYWORDS    %%%%%%%%%%%%%
\medskip{\bf Keywords:} ISM: evolution --- ISM: molecules --- galaxies: ISM --- galaxies: individual(Large Magellanic Cloud) --- radio lines: ISM --- (galaxies:) Magellanic Clouds
% Please write all keywords in lower case. PASA uses the
% standard list of subject headings adopted by The Astrophysical Journal
% and available from http://www.journals.uchicago.edu/ApJ/keywords_text.html.
% Keywords are separated by em-dashes, i.e. ---

%%%%%%%%DO NOT EDIT%%%%%%%%%%%%
\medskip
\medskip
\end{minipage}
\end{changemargin}
]
\small
%%%%%%%%EDIT FROM HERE%%%%%%%%%%%%

%\section{Introduction}
%Please see the PASA Style Guide for help with correct layout for your manuscript.
%Examples of tables and figures are given below.

\section{Introduction}
\label{sec:intro}

The study of star formation (SF) in molecular clouds is an area which
has seen rapid progress in recent years and the Magellanic Clouds are
increasingly seen as one of the most important laboratories for this
research.  At a distance of about 50--60\,kpc and an inclination angle
of $\sim 35^{\circ}$ \citep{vdm01} for the Large Magellanic Cloud
(LMC), current and future instruments can resolve individual
star-forming regions on sub--parsec scales and, at the same time,
assess the global picture on galactic scales. The Magellanic Clouds
also contain a large range of very different environments that can be
observed to test theories of star formation. The 30\,Doradus (30\,Dor)
region in the LMC, for example, is the most vigorous site of current
SF in the Local Group. Consequently, the surroundings of 30\,Dor are
dominated by a strong far--UV radiation field. Possible triggering of
SF on expanding shells can be studied, e.g., in the LMC\,4 region
toward the northeast of the LMC \citep[e.g.][]{bra97,yam01,coh03}. In
addition, the LMC and the Small Magellanic Cloud (SMC) are
gravitationally interacting with each other and with the Galaxy
\citep[cf. the numerical models of, e.g.,
][]{mur80,gar94,sta03,con06}, and the motion of the LMC through the
Galactic halo causes strong ram pressure effects
\citep[e.g.][]{deb98}. It should be noted, however, that tidal or ram
pressure models alone cannot explain the full morphology and dynamics
of the gas and stars in the Magellanic System, and a synthesis of both
is likely required \citep{mas05}. All of these external
interactions, as well as factors internal to the LMC (e.g., shocks,
density waves, and bar-induced motions), influence the ability to form
stars on local and global scales and can potentially be studied in the
Magellanic Clouds in great detail.

Also, with metallicities of $\sim 30$\% and $\sim 10$\% solar in the
LMC and SMC, respectively, cooling and chemistry of gas in the
Magellanic Clouds are clearly different than in the Galaxy and the
conditions may resemble those at higher redshifts, at times when
galaxies were generally less evolved than today. Another important
property of the Magellanic Clouds and in particular the LMC is that
the numbers of resolved giant molecular clouds with associated star
formation is large. For example, \citet{fuk99} counted 107 molecular
complexes in the 2.6$'$ (=37\,pc) resolution NANTEN \co\ survey of the
LMC. It is therefore possible to derive statistically meaningful
results for a large sample of regions, all at the same distance and
therefore resolution.

The location of 30\,Dor and the surrounding massive atomic and
molecular gas reservoir dominating the south-eastern part of the LMC
had led to many speculations on its structure and star forming
history. \citet{fuj90}, for example and, more recently, \citet{bek07}
were able to model this region based on dynamical and hydro--dynamical
simulations of the LMC--SMC interaction, not including the
Galaxy. \citet{gar98}, on the other hand attempted to simulate the
amount of star formation based on the dynamics of the off-center bar
in the LMC. Their simulations, though not designed to predict the
gas distribution, do show that asymmetries can be induced by
the bar. \citet{deb98} pointed out that the ages of structures in the
LMC increase clockwise from the eastern 30\,Dor region toward the
north. They argued that the LMC may undergo SF triggered by a
bow-shock at the eastern edge, where the ram pressure due to the LMC's motion
through the Galactic halo is expected to be greatest.
The clockwise rotation of the LMC then turns the newly
formed, but now aging stellar populations toward the north. In other
words, ram pressure creates a 'hot spot' of SF (currently
30\,Dor) that remains in the eastern part of the LMC while the galaxy
rotation brings fresh gas toward and newly formed stars away from it. If
this interpretation is correct, one may infer that conditions
upstream from 30\,Dor, i.e., from the east
toward the south, may be those encountered just before stars
eventually form. This would be a very fortunate situation as it is
usually difficult to predict the timescales and locations of
future SF in galaxies.

It is common knowledge that stars form out of molecular gas, and
indeed the region south of 30\,Dor contains a remarkably straight,
elongated CO structure which we refer to as the ``molecular ridge.''
\co\ maps from the Columbia survey \citep{coh88}, the NANTEN telescope
\citep{fuk99}, and the SEST key project \citep{kut97,joh98} show that,
with a mass of $10^{7}$\,\msun, about 1/3 of all of the molecular gas
that is traced by CO is found in this structure stretching $\sim
2$\,kpc\ south of 30\,Dor. \citet{joh98} provides an analysis of the
N\,159 and 30\,Dor regions and they find kinetic temperatures in the
10-50\,K range. \citet{kut97} were able to describe the southern part
of the molecular ridge but they restricted their analysis to a
description of the general morphology of the feature and a comparison
of line strength to those in Galactic clouds.

In this paper we present first results of a comprehensive,
high angular resolution \co\ survey of the entire molecular
ridge obtained with the ATNF Mopra\footnote{The Mopra radio telescope
  is part of the Australia Telescope which is funded by the
  Commonwealth of Australia for operation as a National Facility
  managed by CSIRO.} telescope. The observations are the pilot for
the MAGMA project (`The Magellanic Mopra Assessment') which is
currently underway and which will provide high resolution \co\ maps of
molecular clouds across the {\it
  entire} LMC and SMC, mapping all regions where NANTEN has
detected emission, but with a beam size $\sim 16$ times smaller in area. In
terms of sensitivity and resolution, MAGMA is comparable to the SEST
key program but MAGMA encompasses \emph{all} CO traced molecular
clouds rather than a selected ensemble. In Sect.\,\ref{sec:obs} we
describe the observations and data reduction, which is followed by the
presentation of our results in Sect.\,\ref{sec:results}. A discussion
of the results is provided in Sect.\,\ref{sec:discuss} and a summary
in Sect.\,\ref{sec:summary}.

\section{Observations and Data Reduction}
\label{sec:obs}
\begin{figure}[h]
\begin{center}
\includegraphics[scale=0.7, angle=0]{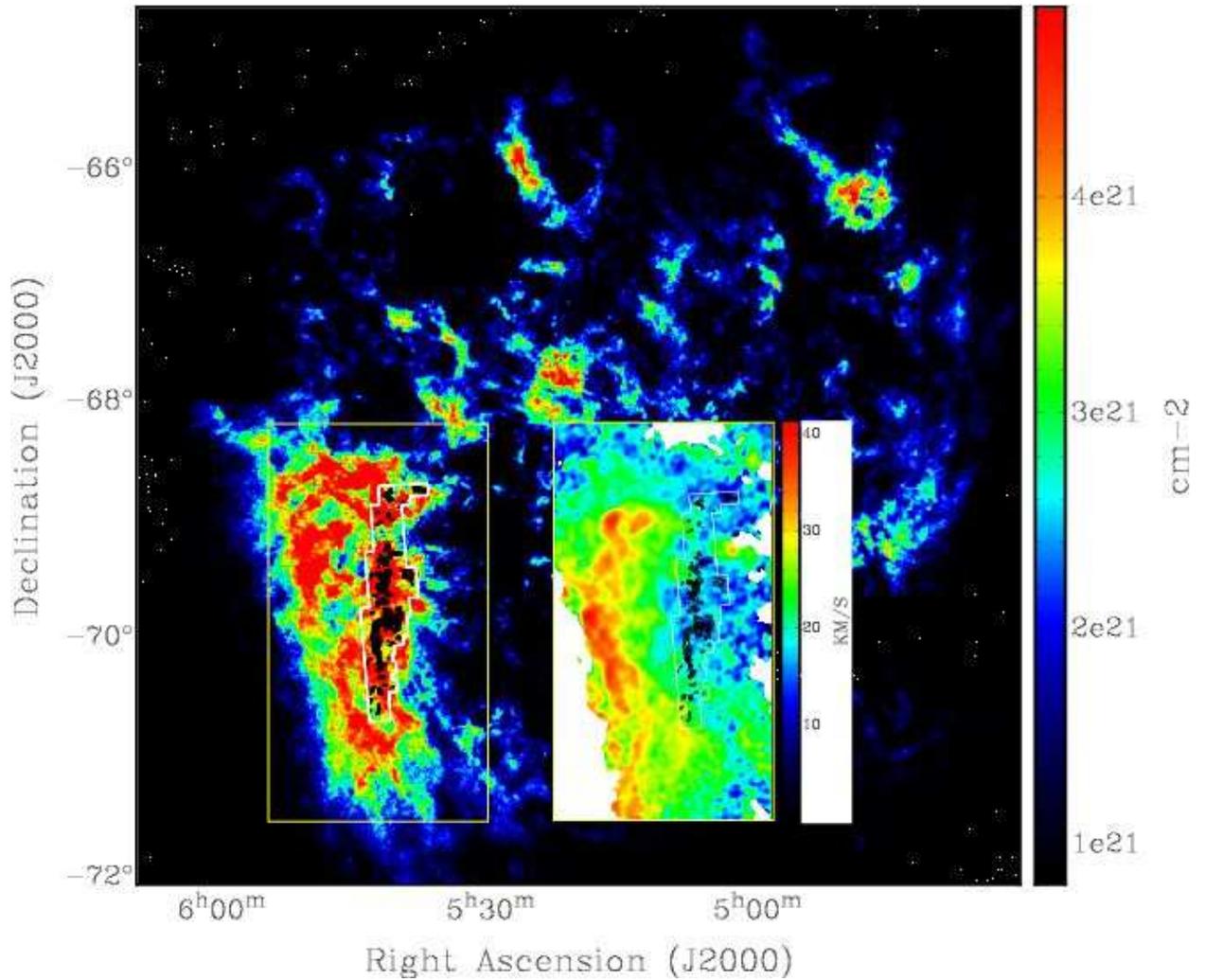}
\caption{\hi\ column density map of the LMC \citep[taken from][beam
  size: 1$'$]{kim98} with the contours of our molecular ridge CO data
  overlaid in black. The white polygon outlines the region
  observed. The yellow boxed insert is the same region as the one
  marked on the column density map but it displays the second moment
  of the \hi\ data (for better signal--to noise, this map was smoothed
  to three times the resolution of the integrated \hi\ map.). Note
  that toward the south-east the second moment is dominated by the
  velocity difference of the disk and 'L' components rather than by
  the line widths of the neutral gas. }\label{fig:global}
\end{center}
\end{figure}

During the 2005 southern winter season, we observed the molecular
ridge close to 30\,Dor with the single dish Mopra telescope in the
\co\ molecular transition (rest frequency: 115.271\,GHz). Mopra was
used in the on--the--fly mapping mode and we observed about $\sim120$
$5'\times5'$ maps in two orthogonal scanning directions with 1.8\,h
integration time each. The total time on source was therefore of order
250\,h. The Mopra AT correlator was set up to observe at a bandwidth
of 64\,MHz split into 1024 channels. This corresponds to a velocity
coverage of $\sim 166$\,\kms\ and a resolution of $\sim 0.16$\,\kms\
and the spectral window was centered on a velocity of 230\,\kms\
LSR. The observation of each map was preceded by pointing calibration
on the bright SiO maser R Dor, with typical corrections less than
5''. $T_{\rm A}^{*}$ system temperatures of the observations were
usually $\sim 600$\,K and the data were calibrated against a warm
absorber that was inserted every $\sim 30$ minutes. The final data
set has a size of $\sim 0.6^{\circ}\times2.0^{\circ}$ which, at an
assumed distance of the LMC of 50\,kpc, corresponds to
$0.5$\,kpc$\times1.8$\,kpc, the long side along declination. A map of
the survey region is shown in Fig.\,\ref{fig:global}. The natural
spatial resolution of Mopra at $\sim 115$\,GHz is about $35"$. The
data was reduced with the ATNF package {\it livedata} to derive the
quotient against line--free reference positions up to $\sim
0.5^{\circ}$ away from the observed fields.  First--order baselines
fitted to line--free channels were subtracted from the resulting
spectra. All data were regridded to produce a
position-position-velocity 3-dimensional data cube using {\it
  gridzilla}. The `inner' Mopra error beam is in the range of
40''-80'' which is similar to the size of the emission and for the
analysis here we applied the extended beam efficiency of $\eta_{\rm
  xb}=0.55$ \citep{lad05} to convert $T_{\rm A}^{*}$ to extended beam
brightnesses $T_{\rm xb}$. For better signal--to--noise ratios, the
data were eventually smoothed to a resolution of $45"$, which
corresponds to a physical length of $\sim 11$\,pc. The rms noise of
the data is about 0.4 K in a 0.16\,\kms\ velocity channel. A
comparison with earlier SEST data at a similar resolution
\citep{joh98} toward the molecular gas in N\,159 shows that the Mopra
and SEST fluxes agree within 10\%.

%Mopra resolution: ~35-40” @ 12CO(1-0)
%(cf. SEST: 43”)
%-> about 9pc resolution at distance of LMC
%
%Size of the map: 2x0.6 degrees = 1.8x0.5kpc  
%
%~120 on-the-fly maps 5’x5’ -> 250h integration time
%
%Using NANTEN data as a ‘cloud finder’  

\section{Results}
\label{sec:results}
\begin{figure}[h]
\begin{center}
\includegraphics[scale=0.4, angle=0]{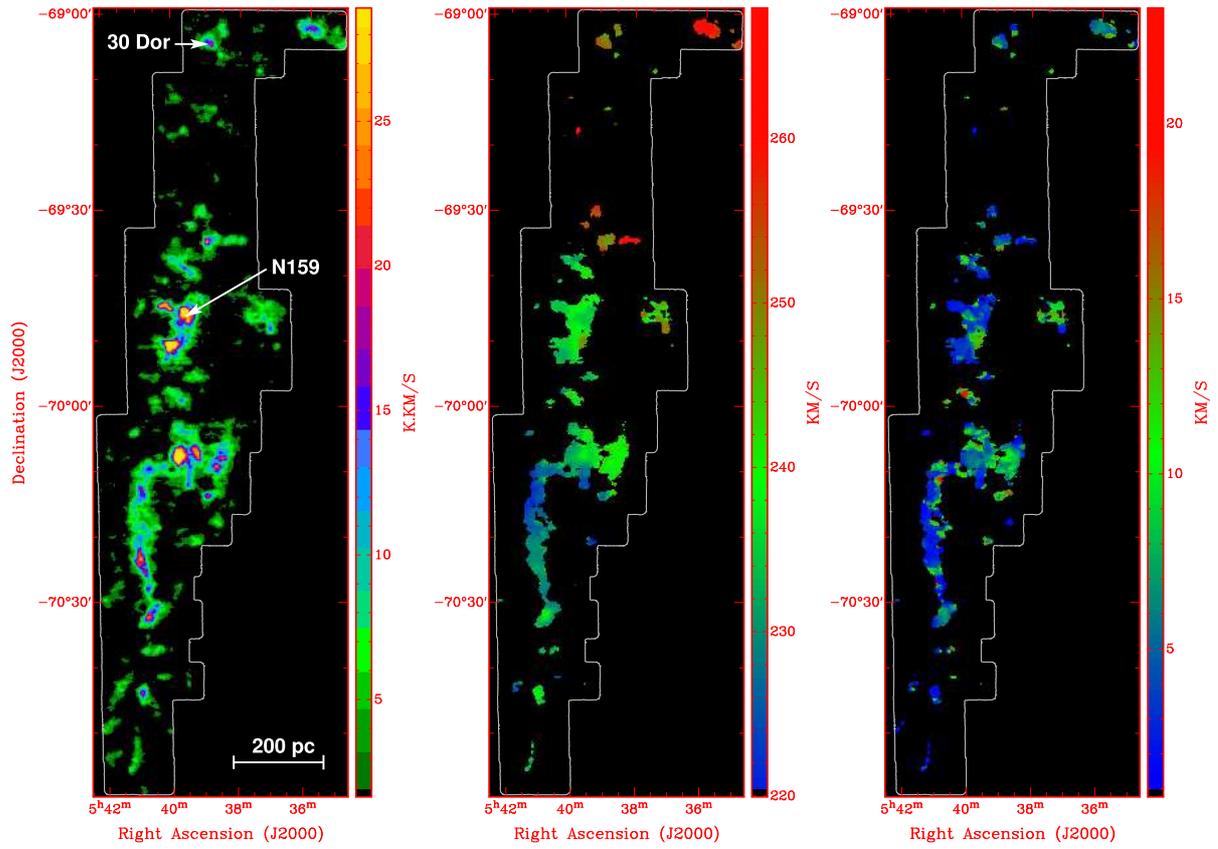}
\caption{Mopra \co\ maps of the molecular ridge close to 30\,Dor. From
  left to right: Integrated intensity in K\,\kms, intensity--weighted
  velocities (moment 1), dispersions (moment
  2). }\label{fig:moments}
\end{center}
\end{figure}

Maps of the molecular ridge derived from the \co\ data cube are
displayed in Fig.\,\ref{fig:moments}. The integrated intensity map
clearly shows that the molecular ridge is a slim (width: $\sim
100-200$\,pc) but long ($\sim 1.8$\,kpc) feature along the declination
axis. The brightest clumps are toward the N159 region (see
Fig.\,\ref{fig:moments}) and about half a degree south of it at the
`kink' of the ridge. Between N159 and 30\,Dor relatively little
molecular gas is observed. Using the \citet{str88} `standard' Galactic
CO intensity--to--\mh\ column density (\xco) conversion factor of
$2.3\times10^{20}$\,cm$^{-2}$ (K\,\kms)$^{-1}$ the detection limit (3
consecutive channels with 3$\sigma$ detection each) is at about
$1.3\times 10^{20}$\,\cden\ (corresponding to a mass of $\sim
200$\,\msun\ within a beam) and for the molecular ridge we derive a
total molecular gas mass of about $3.2\times 10^{6}$\,\msun. But note
that studies of CO with the Columbia 1.2\,m, the SEST and NANTEN
telescopes have shown that the LMC \xco\ factor may be a few times
larger than this Galactic value \citep[][but see Pineda et
al. 2008]{coh88,isr97,fuk99}. For their NANTEN survey data,
\citet{fuk99} use \xco\ of $9\times 10^{20}$\,N$_{\rm H_{2}}$
(K\,\kms)$^{-1}$ and they derive a value of $4\times 10^{7}$\,\msun\
for the entire molecular gas in the LMC. If there are no variations of
\xco\ within the LMC, the ratio of the fluxes should equal the ratio
of masses and if we compare our values to those of NANTEN, we confirm
that the molecular ridge contains about $\sim 1/3$ of the total
molecular mass of the LMC.

\section{Discussion}
\label{sec:discuss}
\begin{figure}[h]
\begin{center}
\includegraphics[scale=0.5, angle=0]{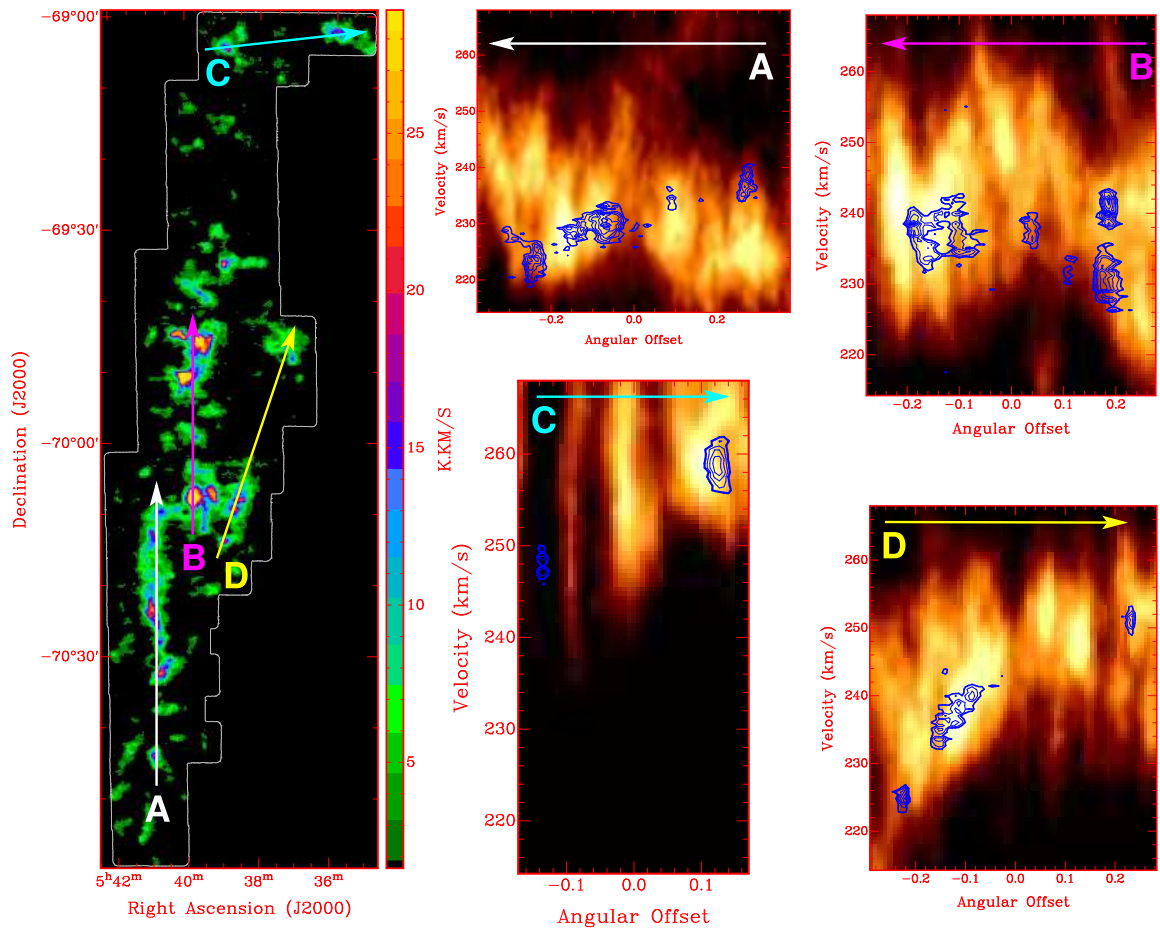}
\caption{Position velocity cuts through the CO and \hi\ data
  cubes. The arrows in the left panel indicate the position,
  direction, and length where the cuts are taken. The corresponding
  position-velocity plots are shown in the middle and to the
  right. Arrows of the same color belong together; they are also coded
  by letters 'A'-'D'. The color scale are data from the \hi\ cube and the
  blue contours are the CO data. Note that the CO at the 30\,Dor
  region exhibits \hi\ in absorption (start of cut 'C').}\label{fig:pV}
\end{center}
\end{figure}

The very thin and long shape of the molecular ridge would argue for it
to be a coherent structure. A comparison of the \hi\ and CO velocities
in various position--velocity diagrams along the ridge
\citep[Fig.\,\ref{fig:pV}, ATCA\footnote{The Australia Telescope
  Compact Array is part of the Australia Telescope which is funded by
  the Commonwealth of Australia for operation as a National Facility
  managed by CSIRO.} \hi\ observations are taken from][]{kim98} shows,
however, that the molecular clumps are not confined to a specific
velocity component of the atomic gas. In addition, the molecular
clumps are somewhat separated from each other, only connected by some
faint common molecular envelope. This is even more obvious when the
3-dimensional position-position-velocity data cube is rotated around
the declination axis. In a more global context the molecular gas in
the ridge finds itself on the western edge of the south-eastern
high-column density region in the LMC, at a position where the \hi\
column densities are still high. However, the \hi\ emission is
bifurcated in two velocity components, called the disk and `L'
component \citep{luk92}. The velocity separation of the two components
is very large towards the eastern edge of the LMC, but they start to
merge into a single velocity component at about the position of the
molecular ridge (see the second moment image in
Fig.\,\ref{fig:global}). This suggests that the clumps in the
molecular ridge were created by the same large-scale triggering event
and thus have similar formation histories, not only in environmental
terms but also on similar timescales.

\subsection{Molecular Cloud Formation}
\label{sec:mcform}

\begin{figure}[h]
\begin{center}
\includegraphics[scale=0.7, angle=0]{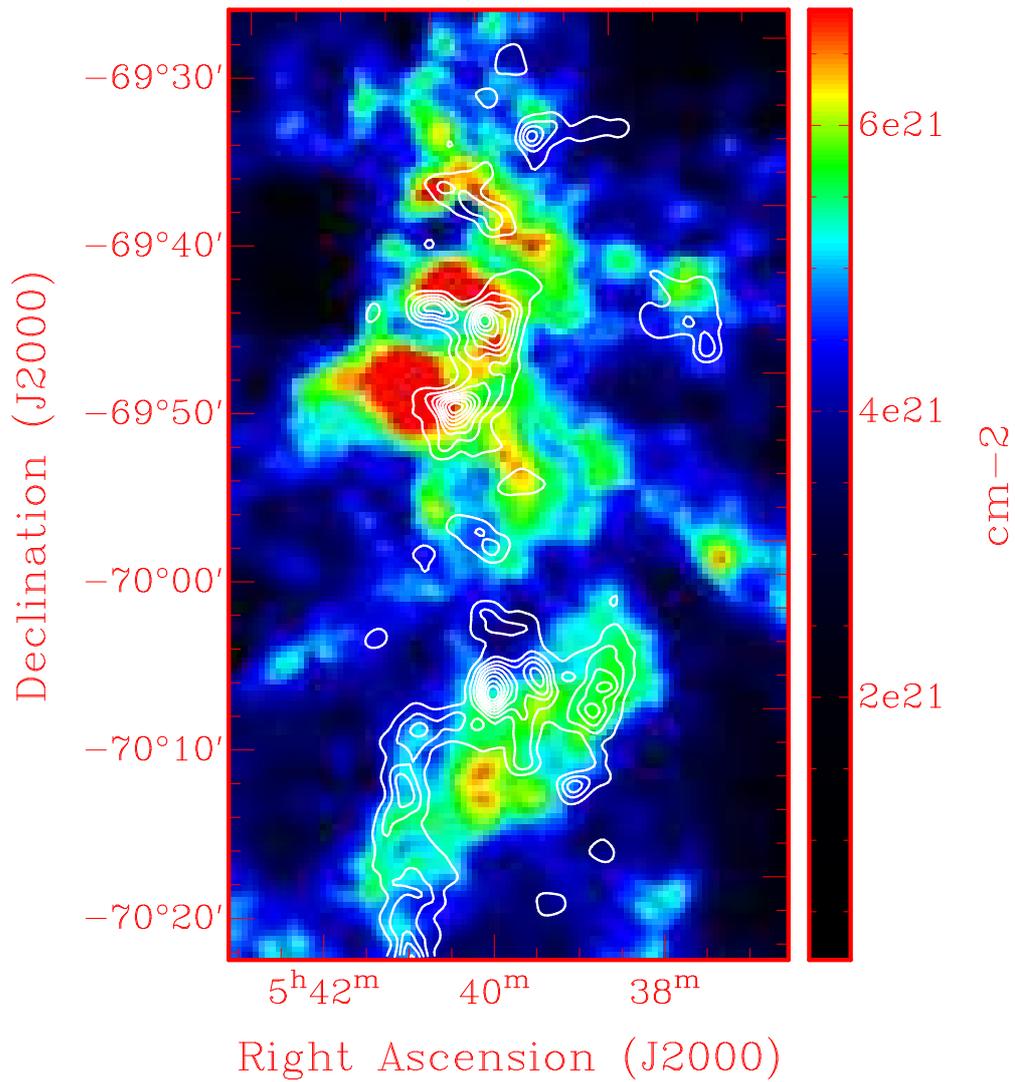}
\caption{CO contours of the region close to N159 overlaid as contours
  on an integrated \hi\ column density map. Note that the CO peaks are
  close to but never coincident with nearby peaks of \hi.}\label{fig:hico}
\end{center}
\end{figure}

\begin{figure}[h]
\begin{center}
\includegraphics[scale=0.3, angle=0]{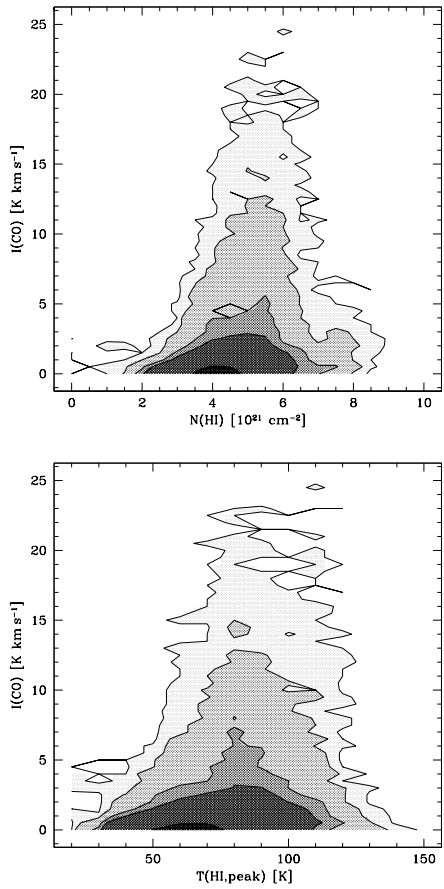}
\caption{Pixel--based scatter plots of the \co\ luminosity as a function
  of \hi\ column density (top) and \hi\ peak brightness (bottom). The
  contours are levels of constant point density in logarithmic
  spacings. }\label{fig:scatter}
\end{center}
\end{figure}

In Fig.\,\ref{fig:hico}, CO contours are overlaid over an integrated
\hi\ column density map. The CO peaks are usually found very close to
\hi\ peaks but they virtually never coincide. Such a morphology was
already indicated by the lower resolution NANTEN data (resolution:
$\sim 2.6'$) and with the high angular resolution of Mopra, this is
unambiguously confirmed. If, instead of the integrated \hi\ column
density, the \hi\ peak flux is compared to the \co\ intensity, the
peaks of the two gas tracers are getting slightly closer to each other
but they still never coincide. To quantify this effect, pixel-to-pixel
correlations between the CO luminosity and the \hi\ column density and
\hi\ peak brightness temperature are shown in
Fig.\,\ref{fig:scatter}. The CO bright pixels are at intermediate \hi\
column densities and brightnesses and not at the largest values. In
particular, the \hi\ peak brightness values where the brightest CO
emission is observed hovers around 60 to 100\,K. In the following, we
will discuss different mechanisms that may be responsible for the
shift of the CO and \hi\ emission peaks:

\paragraph {\it 1. Rapid conversion of warm \hi\ into \mh.} At a certain
threshold, the atomic gas combines into molecular hydrogen and more
than just the gas above the threshold is converted, creating slight
depressions in the map of atomic \hi. If the \hi\ gas remains
optically thin one can determine the entire amount and distribution of
hydrogen atoms from both, \hi\ and \mh\ in the gas. In this scenario
one would expect a smooth distribution of hydrogen, some in atomic,
and some in molecular form. As an example we consider N159. The
atomic gas in this region has typical column densities of $\sim
7\times10^{21}$\,\cden\ whereas the \hi\ at the positions of the CO
peaks have a somewhat lower column of N(\hi)$\sim
5\times10^{21}$\,\cden. The CO peaks (smoothed to the resolution of
the \hi\ map, 60$''$) have intensities of $\sim
35$\,K\,\kms. Converted with a Galactic $X_{\rm CO}$ factor, this
becomes an \mh\ column density of N(\mh)$\sim 8\times10^{21}$\,\cden\
which adds to the aforementioned \hi\ column density of $\sim
5\times10^{21}$\,\cden. In total, the hydrogen column density at the
positions of the molecular clouds is determined to be N(H$_{\rm
  total}$)=N(\hi)+2\,N(\mh) $\approx 21\times10^{21}$\,\cden. This is a
factor of $\sim 3$ larger than the \hi\ column that surrounds the
molecular clumps but where no CO is detected. Therefore, the total
(\hi\ + 2 \mh) hydrogen column density map exhibits a jump rather than
a smooth transition whenever molecular gas is present. This sharp edge
cannot be softened by a variation of the \xco\ factor. E.g.,
\citet{fuk99} and \citet{isr97} find that the LMC has an \xco\ factor
a few times larger than the Galactic value. Using such a higher \xco\
would result in larger \mh\ columns and therefore in an even steeper
rise of total hydrogen column densities. If the \xco\ factor is
changed to produce smooth hydrogen maps, where the \mh\ traced by CO
fills in the 'holes' of \hi, one requires an \xco\ of about 3 times
\emph{less} than the Galactic value. No study of the LMC or of any
other low metallicity object has shown such low \xco\ factors
before. 

This comparison, however, is only valid if all molecular gas is traced
by CO. If there are extended regions with molecular gas that is not
traced by CO, e.g., at low densities when CO may be dissociated by the
surrounding radiation field but \mh\ is not, the sharp contrast in the
total hydrogen map may be smoothed out. Such an \mh\ phase is hard to
detect and one of the best approaches is UV absorption
spectroscopy. \citet{tum02} performed such a survey toward many
sightlines in the LMC with {\it FUSE}. They find that the molecular
hydrogen, with a fraction of $\sim 1$\% (cf. Galaxy: $\sim 10$\%), is
not a huge contributor to the entire gaseous ISM \citep[see also the
{\it ORFEUS} UV absorption observation presented in][]{deb98b}. They
were also observing sightlines close to 30\,Doradus where the UV
radiation is very strong. If there are large amounts of \mh\ that are
not traced by CO due to the higher dissociation energy of \mh\, they
should be most obvious toward that region. Indeed, the total column of
\mh\ they find close to 30\,Dor is with $\sim 10^{20}$\,\cden\ one of
the largest in their survey. However, the \hi\ column density towards
the same sightline is $\sim 70$ times larger. Similar fractions are
found all across the LMC and the column of \mh\ in absorption never
exceeds that of \hi. It is therefore unlikely that the contrast
between the \hi\ and \mh\ column densities as traced by CO could be
smoothed out with extended layers of molecular gas that are undetected
in CO observations. This is in contrast to dust extinction studies
which suggest that there may be large, not CO traced reservoirs of
molecular gas in the LMC \citep[e.g.][]{ima07}.

To conclude, the shifts of the \hi\ and the CO peaks are probably not
simply a result of warm hydrogen atoms (WNM) combining into
\mh. However, gravitation may accumulate a lot of material quickly in
the densest molecular cores. Typical shifts between CO and \hi\ peaks
are of order 2$'$, which corresponds to $\sim 30$\,pc. Within a
typical free fall time of a Myr, self-gravitation would accelerate the
gas to a velocity of about 30\,\kms, a velocity gradient that is not
ruled out by the position velocity diagrams displayed in
Fig.\,\ref{fig:pV}.

%HI -> H2 conversion 
%  but: proton map shows HI <> H2 phase jump
%  Maybe simult. conversion + self-gravitation 

\paragraph {\it 2. WNM to CNM to \mh\ conversion.} 
As shown in Fig.\,\ref{fig:scatter}, the \hi\ brightness temperature,
at the location where most and the brightest molecular material
resides, falls into the range of 60 to 100\,K which corresponds to the
kinetic temperature of the cold neutral medium in the Galactic disk
\citep[CNM, e.g., ][]{dick90}. A conversion of WNM to CNM can be
achieved by a shock. Indeed, the atomic gas to the east of the
molecular ridge is prone to ram pressure effects as this side is the
leading edge towards the Milky Way halo. The region is also influenced
by tidal effects \citep[e.g.][]{sta03,mas05} and it has even been
speculated that this region is responsible for the origin of the
Magellanic Stream \citep{nid07}. The second moment map of the \hi\ in
Fig.\,\ref{fig:global} is in fact not dominated by the intrinsic line
width of the neutral gas but by the separation of disk and 'L'
velocity components. At the eastern edge the separation is very large
and the two components become closer in velocity space toward the
western direction. At the position of the molecular ridge the disk and
'L' components start to merge. To date the origin of the bifurcation
is still under discussion \citep[e.g.][]{bek07,nid07} but whether it
is ram pressure or tidal effects or both, the shear and shocks from
such an interaction introduces energy into the ISM that may reduce the
likeliness to form molecular clouds. Indeed, some molecular complexes,
the 'arc' clouds, are found in the region \citep[][]{miz01} but the
bulk of molecular gas, the molecular ridge, is found where the disk
and 'L' components start to merge into a single velocity component.
If this indicates that they are also spatially merging, one may
speculate that only the combination of both components can provide
enough shielding for the gas to cool down and to become molecular.

In other words, shocks and/or shearing forces propagating from the
eastern edge may convert some of the WNM into CNM. When the disk
and 'L' component merge they may provide enough shielding to convert
part of the CNM into molecular gas and the remaining, cold \hi\
becomes optically thick and thus levels off at a brightness temperature
close to the kinetic temperature of the CNM.

\paragraph {\it 3. \hi\ accretion.} 
The above process does not have to be static. As discussed earlier, we
cannot rule out gas moving at speeds that are required to match
free-fall timescales. \hi\ accretion after the formation of the first
molecular cores may be responsible for the pile--up of neutral gas
close to the molecular clouds. In this scenario, once the atomic gas
comes close to the molecular clumps, the gas is shielded, cooled, and
converted into \mh\ \citep[see also][]{fuk07}. \hi\ accretion may
therefore be described as a cooling flow. Cooling flows would result
in decreasing \hi\ line widths toward the molecular cores. We do not
observe this on the scales of the \hi-CO peak offsets but the most
significant cooling may happen only very close to the molecular cores,
too close to be resolved with the ATCA \hi\ observations at 21\,cm.

\begin{figure}[h]
\begin{center}
\includegraphics[scale=0.4, angle=0]{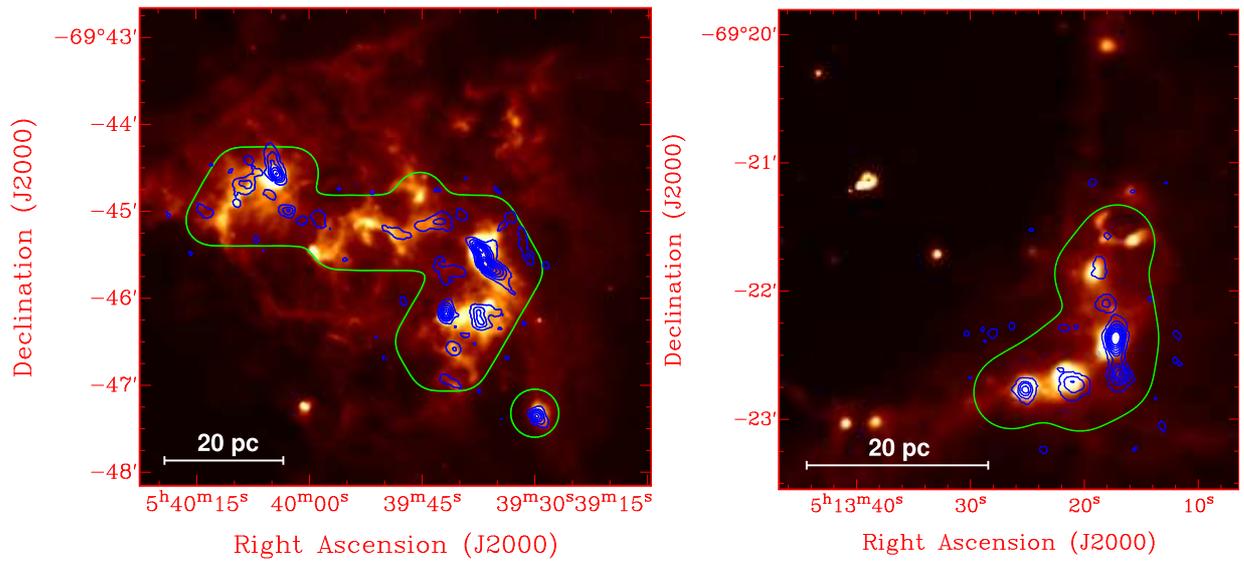}
\caption{The star-forming regions N159 (left) and N113 (right) in the
  LMC. The contours show the integrated HCO$^{+}$ emission as observed
  with the ATCA on top of {\it Spitzer} 8$\mu$m maps. The green lines
  outline the mapped regions.}\label{fig:hco}
\end{center}
\end{figure}

\paragraph {\it 4. Mechanical feedback from massive stars.}
Newly formed, massive stars deposit large amounts of energy into the
surrounding interstellar medium (ISM) in the form of strong stellar
winds and supernovae. Virtually all star-forming galaxies exhibit an
ISM that is dominated by a wealth of expanding shells of all
sizes. \citet{kim07} catalog the expanding shells in the LMC. Along
the molecular ridge, the number density of the shells seems to be
lower than in the rest of the LMC. It should therefore not have a
significant effect on the distribution of the atomic and molecular
gas. For some regions like the N\,159 region, however, some
shell--like structure is observed in \hi\ and dust maps (cf.
Fig.\,\ref{fig:hco}) and the CO emission toward this position may
coincide with the inner side of the ring, similar to what is
observed toward other expanding Galactic shells or LMC\,4 \citep{yam01,cap05}. \\

The scenario that best fits the observations is the second one, in
which shocks transform WNM into optically thick CNM followed by
conversion into \mh, with or without \hi\ accretion. The very sharp
rise of the total hydrogen column density inferred at the position of
the molecular gas is likely an artifact of high \hi\ optical depth.
On the other hand, the confinement of CO emission to clumpy structures
suggests an important role for self-gravity in shaping the molecular
clouds.

\subsection{Star Formation}
\label{sec:starform}

Typical \hi\ column density thresholds in massive and dwarf galaxies,
above which SF is observed hover around a canonical value of $\sim
1\times10^{21}$\,\cden\ \citep[e.g.][]{wal07}. Below this threshold
SF appears to be suppressed.  For most of the LMC this rule holds
well. However, the south-eastern region, dominated by ram pressure and
tidal forces of the LMC/SMC/Milky Way interaction, appears to be
different. Virtually no SF, as traced by \ha, is observed in this
region at positions where the \hi\ column density is below
$5\times10^{21}$\,\cden. This may be due to the tidal and ram pressure
forces in the two gas components of the south--eastern LMC.

On smaller scales, SF occurs in the very densest molecular clumps. We
conducted a survey with the ATCA of the high-density gas tracers
HCO$^{+}$ and HCN towards the star-forming regions N159 (40 pointing
mosaic) in the molecular ridge and toward N113 (7 pointing mosaic) at
the western tip of the LMC bar. The observations have a resolution of
$\sim 7''$ which, at the distance of the LMC, corresponds to a
physical length of $\sim 1.7$\,pc. Both of the observed molecules are
only excited at volume densities exceeding $\sim
10^{4}$\,\vden. \hcop, however, is also a tracer for photo-dominated
regions and can also form in more diffuse material where the C$^{+}$
abundance is high. Integrated HCO$^{+}$ intensity maps overlaid on
8$\mu m$ Spitzer maps are shown in Fig.\,\ref{fig:hco}. In both
regions, the dense molecular gas typically overlaps or is adjacent to
bright infrared (IR) emission peaks. These peaks are the locations
where individual stars and stellar clusters eventually form within the
larger molecular complexes traced by CO. Comparing the HCN and \hcop\
maps, we find that, after beam deconvolution, the \hcop\ emission is
about 20-30\% more extended than the HCN clumps, in agreement with
the results presented in \citet{won06}. The $8\mu$m IR morphology of
N159 appears to be ring--like and indeed it is a 1-2 Myr old
wind--blown bubble created by a central, massive star
\citep{joh05}. Most of the dense molecular gas is situated at the rim
of this shell, with relatively little observed at projected sightlines
towards the interior. N113, in contrast, is a long, bent string of
several individual knots, both visible in the IR and in the HCO$^{+}$
and HCN maps. The string of knots curves from the south-east towards
the north. This morphology appears reminiscent of a circular arc,
and close to the center of this hypothetical circle, towards the
north--east, one indeed finds HD\,269219, a $\sim 30$\,\msun\
supergiant B star and HD\,269217, an emission-line star. Thus, the
densest pockets of molecular gas and the associated star-forming
regions appear to be located on a shock front created by the two
energy--injecting massive stars, not unlike the situation of the shell
in N159.

The two examples, and the above discussion on molecular cloud
formation argues that shocks are the major mechanism to trigger both,
the formation of low-density molecular clouds and the formation of
star--forming, high-density molecular clumps within these larger
structures.

\section{Summary}
\label{sec:summary}

The Large Magellanic Cloud is a unique object to study molecular cloud
and star formation on galaxy-wide scales with high spatial detail and
in a large range of very different environments. A very prominent
molecular feature stretches from 30\,Dor $\sim 1.8$\,kpc southward; it
contains about 1/3 of the entire molecular content of the LMC, as
traced by CO. Our Mopra data of this molecular ridge reveals the
following:

\begin{itemize}
\item The structure is most likely not a coherent, large molecular
  complex but consists of smaller molecular clouds that are observed
  to be embedded in different velocity components of the neutral
  atomic gas. However, the global picture of the LMC shows that the
  molecular ridge is at the innermost edge of the south--eastern
  high-density region of the LMC, a region where tidal and ram
  pressure forces have a major influence on the state of the gas. The
  ridge coincides with \hi\ at high column density and is just where
  the disk and 'L' \hi\ velocity components start to merge. Thus, the
  molecular clouds in the ridge are likely to share the same formation
  histories and timescales.

\item The local \hi\ and CO peaks are typically displaced by $\sim
  30$\,pc. They are usually near each other but hardly ever
  coincident. A likely explanation for this morphology is that shocks
  propagate from the eastern edge of the LMC and turn warm \hi\ into
  partly optically thick, cold \hi\ and eventually into molecular
  hydrogen. This may happen in a static sense or via \hi\
  accretion. The hydrogen column densities within the molecular clumps
  are much larger than in the surrounding material, which indicates
  that the clouds are self--gravitating. Other mechanisms, such as
  feedback from massive stars may also be responsible for the shift of
  \hi\ and CO peaks but are likely only playing a role for a small
  number of molecular complexes.

\item Very dense pockets of molecular gas, as traced by HCO$^{+}$ and
  HCN, are found near and coinciding with strong IR emission. This
  strengthens the suggested correlation of the very dense gas and
  actual star formation rates. For the N159 and N113 regions we find
  the high density gas peaks to be mostly located on shock fronts and
  on the rims of expanding shells, powered by central, massive stars.

\end{itemize}

Over the coming years, the study of molecular gas in the LMC and SMC
enters a new era. Our team will continue to map all the molecular gas
in both systems with the Mopra (MAGMA survey) and ATCA telescopes. In
addition, many new facilities are currently under construction or have
just been completed that can be used to observe the LMC, such as APEX,
ASTE, NANTEN2, ALMA, as well as satellites such as {\it Herschel} and
{\it Spitzer}. Studying molecular gas in the LMC provides a unique
link between Galactic and extragalactic observations and will be
germane to the full understanding of star formation on all scales.

%% It is preferable to embed your figures in the text as in the following example
%\begin{figure}[h]
%\begin{center}
%%\includegraphics[scale=1, angle=0]{figure.eps}
%caption{An example figure caption.}\label{figexample}
%\end{center}
%end{figure}

%
%
%%%Format tables as in the following example
%\begin{table}[h]
%\begin{center}
%\caption{Example Table}\label{tableexample}
%\begin{tabular}{lcc}
%\hline Column 1 & Column 2 & Column 3 \\
%\hline Table Content$^a$ \\
%\hline
%\end{tabular}
%\medskip\\
%$^a$Table footnotes go here.\\
%\end{center}
%\end{table}
%
%
%\section*{Acknowledgments} %If needed

\clearpage

\clearpage

\clearpage

\clearpage

\clearpage

\clearpage

%\end{multicols}

\end{document}